# Temperature sensing using nitrogen-vacancy centers with multiple-poly crystal directions based on Zeeman splitting

**Li Xing[1], Xiaojuan Feng[1*], Jintao Zhang[1*], and Zheng Wang[1,2]**

[1]*National Institute of Metrology, Beijing 100029, People's Republic of China*
[2]*Department of Precision Instrument, Tsinghua University, Beijing 100084, China*
*Email: *fengxj@nim.ac.cn*
*Email: *zhangjint@nim.ac.cn*

**ABSTRACT**

We demonstrate a novel method based on the Zeeman splitting of electronic spins to improve the performance for temperature sensing of negatively-charged nitrogen-vacancy (NV) centers in multiple-poly diamond. The theoretical model for selection principle of resonance peaks corresponding to a single NV axis for determining the temperature dependence is clarified. The spectral linewidth is effectively narrowed and the thermometer is insensitive to magnetic field fluctuations. Repeatability and accuracy of the relationship calibration between the zero-field splitting (ZFS) parameter $D$ and temperature $T$ in the range of 298 K to 323 K is significantly improved, and the results of coefficient $dD/dT$ is 75.33 kHz/K. Finally, this method promotes the average temperature measurement sensitivity (below 10 Hz) of our setup from 0.49 K/Hz$^{1/2}$ to 0.22 K/Hz$^{1/2}$.

## 1. INTRODUCTION

Recently, the NV centers in diamond are promising temperature sensors of nano- to milli-meter scales[1-4] applying for chip manufacturing[5], biological cell and tissue structure recognition[6,], due to the sensitive temperature dependent polarization of their ground triplet electronic spin states, good biological compatibility and physical stabilities of diamond[8-11]. The temperature of diamond sensing can be acquired by an optically detected magnetic resonance (ODMR) spectrum, which generates from the NV fluorescence in a continuous-wave (CW) around 2.87 GHz[12,13].

The key to realize precise temperature sensing is to improve the measurement sensitivity and obtain accurate *D-T* relationship. The sensitivity of temperature measurements $\eta$, characterizing the measurement signal-to-noise ratio, is limited by the fraction $\Delta v/C$, where $\Delta v$ denotes the linewidth of resonant profile and *C* represents the contrast of the fluorescence detection[14,15]. Therefore, narrowing spectral linewidth $\Delta v$ and increasing fluorescence contrast *C* are conducive to improve the sensitivity. However, the ground state energy level of the NV centers is also sensitive to external magnetic field[10,16]. In the absence of magnetic shielding system, the linewidth of ODMR spectrum can be widened by the environment residual magnetic fields. Moreover, a large power of microwave is usually used to increase the fluorescence contrast, which will also lead to the wider linewidth[17,18]. Generally, the linewidth can be narrowed by optimizing the power of microwave and laser light, which should measure multiple groups of ODMR spectrum at different powers[19,20]. A. M. Wojciechowski et al. proposed a method for precision temperature sensing by driving two hyperfine transitions of different electronic states of NV centers, and could be immune to the impact of magnetic field[10]. Furthermore, some modulation methods for microwave have been implemented to isolate interferences from undesired physical quantities, which could suppress the magnetic field noise, microwave power shift and so on[5,10,21]. At present, the temperature dependent differences of NV centers among different research groups have not been studied in depth[22-24]. We have briefly discussed the possible reasons for the difference in *D-T* relationship mainly involving the discrepancy of diamond samples and the effect of microwave and laser[17,22,23]. Therefore, it is urgent to put forward more novel methods to improve the measurement performance and calibration accuracy of *D-T* relationship.

In this paper, we demonstrate a method for temperature sensing of NV centers based on the Zeeman splitting to solve the above problems. The temperature dependence of the NV centers over the temperature range of 298 K to 323 K was tested using the Zeeman splitting and ZFS methods, respectively. Comparing with the ZFS method, the spectral linewidth was narrowed and the thermometer was insensitive to magnetic fields. Finally, our approach improved the accuracy of *D-T* relationship calibration and the temperature measurement sensitivity.

## 2. PRINCIPLE

The electronic spins of NV centers can be described by the ground-state Hamiltonian[8,25],

$$\mathcal{H}_s = DS_z^2 + E\left(S_x^2 - S_y^2\right) + \gamma_e \vec{B}\cdot\vec{S}, \tag{1}$$

where, $\gamma_e$ is gyromagnetic ratio of the electronic spin. $\vec{S} = (S_x, S_y, S_z)$ is the spin operator of the electronic spin quantum number for $S=1$. $E$ is the transverse ZFS parameter, which is related to the stress and electric field of diamond lattice[26]. $D$ is the axial ZFS parameter, which is generated by spin-spin interaction of the two unpaired electrons in NV center, and will split the levels between $|m_s = 0\rangle$ and $|m_s = \pm 1\rangle$[12]. $D$ is temperature dependent and can be used to sense temperature variation. $\vec{B}$ is an applied magnetic field, and will cause Zeeman splitting of the degenerate $|m_s = \pm 1\rangle$ levels[21,27]. Therefore, both of the temperature and external magnetic fields have influences on the Hamiltonian of the ground electronic spin states, as shown in Figure 1(a).

The ZFS parameter $D$ is principally dependent on the thermal equilibrium temperatures of the diamond. Thermocouples or thermistors are usually utilized to provide the reference temperature for the calibration of $D$-$T$ relationship[8,17], so that temperature sensing using NV centers is currently considered as a secondary measurement method. In general, the CW-ODMR spectral line shown in Figure 1(b) is used to obtain the value of $D$[8]. In the absence of an applied magnetic field (only ZFS without obvious Zeeman level splitting), the splitting of the resonance peak is due to non-zero $E$, caused by local strain from internal defect of crystal[5]. Thus, the two cumulative Lorentzian lines should be used to fit the double-peak line to obtain parameter $D$, which are difficult to fit and lead to serious fitting errors. Moreover, the structure of NV centers in diamond crystal lattice is $C_{3v}$ symmetry. There are four different NV axes[28,29], and the electronic spins of different NV axes will interfere with each other. Furthermore, the linewidth introduced by the environment residual magnetic fields and inhomogeneous lattice stress need to be narrowed to improve the sensitivity limit.

In the presence of an applied magnetic field, the electronic spins in each NV axis can produce the Zeeman splitting. Its transition frequencies are affected by the projection of magnetic field on the NV axis. When dozens of gauss of static magnetic field along the [111]-oriented diamond is applied, the ODMR spectrum involving two pairs of resonance peaks is shown in Figure 1(c). In this case, the resonances of a single NV axis appears while the attenuation of fluorescence contrast can be possibly prevented. The outside pair of resonance peaks corresponds to the NV centers spin along the [111] direction, and the inner-side pair of resonance peaks corresponds to the other three NV axes. Thus, the outside resonance peaks can be selected to realize precise *D-T* relationship measurement, and the temperature measurement based on Zeeman splitting can also avoid the mutual interference among the electronic spins of different NV axes. Considering the magnetic field and temperature sensing, the resonance frequency of the $|m_s = \pm 1\rangle$ states with the Zeeman splitting can be described by calculating the Hamiltonian's eigenvalue of electronic spins,

$$f_\pm = D \pm \gamma_e B_i, \qquad (2)$$

where $B_i$ represents the projected component of the external magnetic field on the *i*-th NV axis (*i*=1,2,3,4). Thus, when the value of *D* varies with temperature change, all resonance peaks on the ODMR spectrum will move in the same direction, and the resonance frequency shift is $\Delta f = \frac{dD}{dT}\Delta T$. When the applied external magnetic field changes, the Zeeman splitting resonances corresponding to $|m_s = +1\rangle$ and $|m_s = -1\rangle$ will shift in the opposite direction, and the center frequency of the resonant lines is still at the ZFS parameter *D*. The center frequency is $f_e = (f_+ + f_-)/2 = D$, so that the variation of *D* with temperature change can be obtained through the Zeeman splitting of electronic spins. This method can effectively avoid the effect of magnetic field fluctuations on the *D* frequency shift because of $\partial f_+ / \partial B = -\partial f_- / \partial B$. The influence of magnetic field noise on temperature sensing can be significantly suppressed. Moreover, the method based on Zeeman splitting can effectively narrow the linewidth, and suppress the influence of local stain in ODMR spectrum. The resonance peaks can be directly fitted with Lorentz curve, which can reduce the fitting errors. Comparing with the method based on ZFS, the Zeeman splitting method of single NV axis, forming two pairs of resonance peaks, can theoretically improve the calibration accuracy of *D-T* relationship and the temperature measurement sensitivity.

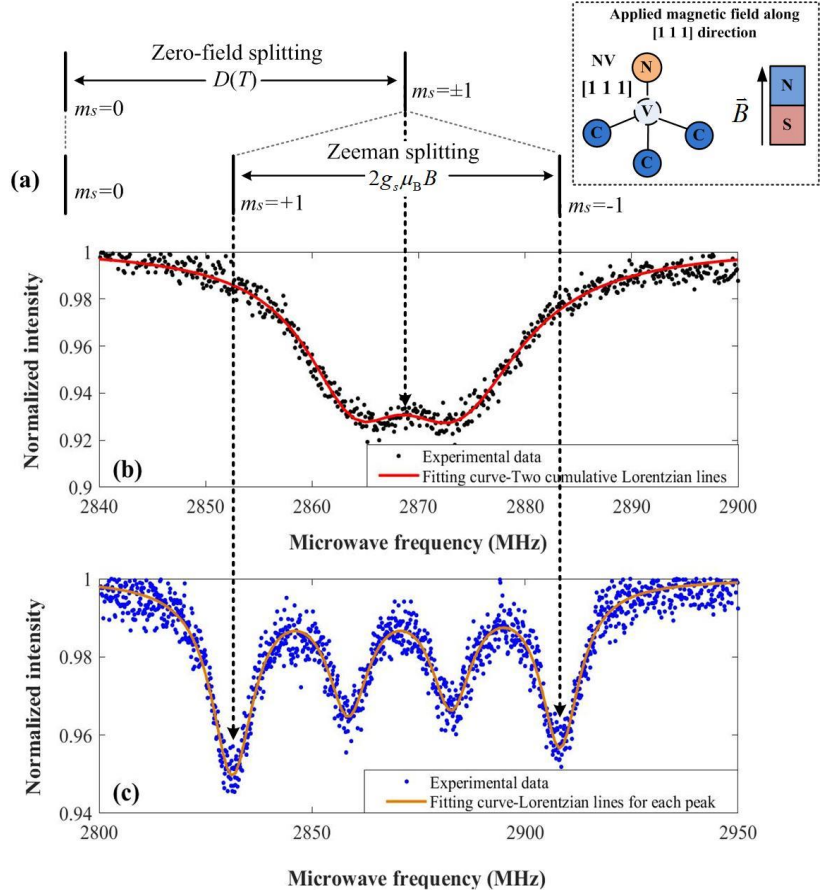

**Figure 1.** Energy levels of electronic spins and ODMR spectrum. (a) NV center ground state levels of electronic spins with ZFS and Zeeman splitting. (b) ODMR spectra without an applied magnetic field, and fitted with a two-cumulative Lorentzian line. (c) ODMR spectra with an [111]-oriented bias magnetic field, and fitted with normal Lorentzian lines.

## 3. EXPERIMENTAL SETUP

The schematic of the setup is shown in Figure 2. The Type-Ib diamond sample of 3 ×3×0.3 mm³ we used were subjected to electron irradiation and annealing, whose concentration of NV centers was larger than 10 ppm. The diamond was glued to a transparent glass-slide, and the glass was fixed on a printed circuit board (PCB). The flexible heating film was pasted on the other side of glass-slide, which could heat the diamond to 323 K. A type K thermocouple with outer diameter of 0.6 mm was attached to the diamond to provide reference temperature values for NV center temperature measurement. The thermocouple would not generate additional interference of thermal effect because it required no current for measurement. A straight wire microwave antenna with diameter of 60 μm was fixed on the diamond surface, and applied the microwave to NV centers. The microwave was generated by a signal generator (Rohde & Schwarz SMIQ04B) with resolution of 0.1 Hz and enlarged by an amplifier (Mini-

Circuits ZHL-16W-43-S+). A pumping light at 532 nm was applied to excite the NV centers in diamond, which was focused on the surface of diamond through an objective (×100/1.25, OLYMPUS). The fluorescence with wavelength of 600 nm-800 nm emitted from the NV⁻ and then was collected by the objective. A dichroic mirror and a long-pass filter (cut on wavelength: 600 nm) were applied to prevent the 532 nm laser from entering the avalanche photodiode (APD), and ensured that the detection of fluorescence would not be interfered. We used a permanent magnet to generate a constant external magnetic field along the [111] direction of diamond to make the Zeeman splitting of electron spins.

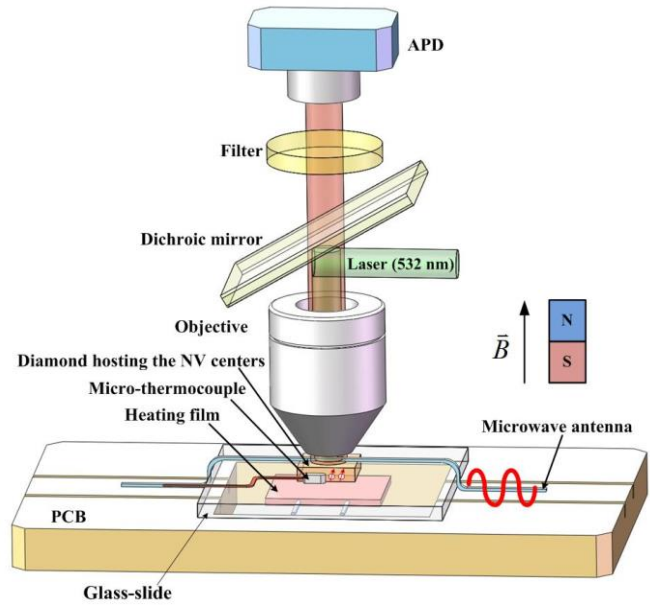

**Figure 2.** The schematic of the experimental setup. APD: avalanche photodiode, PCB: printed circuit board.

## 4. EXPERIMENT RESULT AND DISCUSSION

The relationship between zero-field splitting parameter D and T should be accurately calibrated to realize temperature measurement. A series of ODMR spectra at temperatures ranging from 298 K to 323 K, -5 dBm of microwave power and about 10 mW of pumping laser power have been tested. Firstly, we obtained the values of $D$ from the double-peak structure by two-cumulative Lorentz fitting when there was no external magnetic field applied. The linewidth of the spectra was about 21 MHz. According to Eq. (2), the values of $D$ calculating from the Zeeman level splitting with an applied magnetic field have been obtained by single Lorentz fitting curve, which could avoid the interference among electron spins of different NV axes. The linewidth of a single resonance peak was about 9 MHz. The calibration results of $D$-$T$ are shown in Figure

3, and directly fitted with linear curve. The temperature-induced coefficient d$D$/d$T$ of our system based on Zeeman splitting is 75.33 kHz/K, and the difference is 0.1 kHz/K compared with the ZFS method. Moreover, the standard deviation (STD) of $D$-$T$ measurement results based on Zeeman splitting is much smaller than that of results based on ZFS method, mainly due to the narrowing of the linewidth and the improvement of fitting accuracy for ODMR spectra. When the measurements are carried out using ZFS method, the minimum and maximum STD of $D$ is 70.93 kHz at 298 K and 607.49 kHz at 313 kHz. However, the minimum and maximum STD of $D$ based on Zeeman splitting method is 19.66 kHz at 298 K and 68.10 kHz at 308 K, respectively. Thus, the method using Zeeman splitting can improve the repeatability and accuracy of $D$-$T$ relationship measurement.

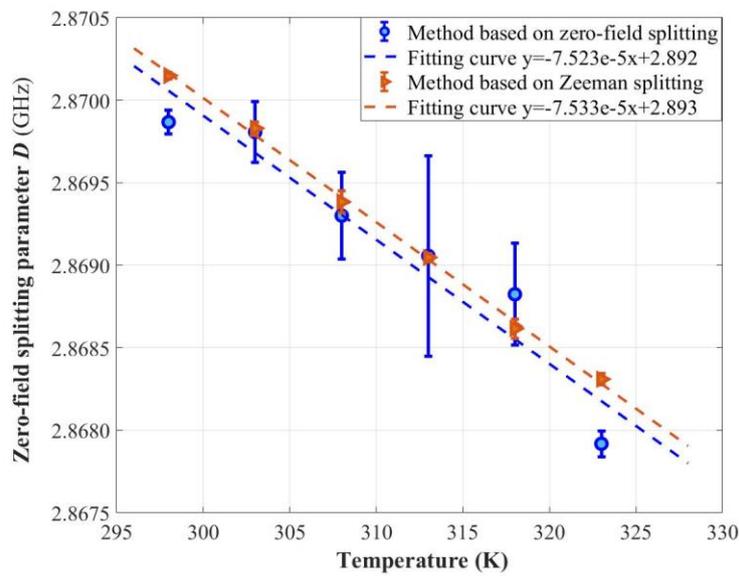

**Figure 3.** The measurement results of $D$-$T$ at temperatures ranging from 298 K to 323 K.

In order to convert the unit of output fluorescent voltage signal to Kelvin, we acquired the scale factor by obtaining the slope of ODMR spectral line where the fluorescence varies greatly with the microwave frequency sweep[30]. The scale factors based on Zeeman splitting (with an applied magnetic field) and zero-field splitting (without an applied magnetic field) are shown in Figure 4. The scale factors for method based on Zeeman splitting at different temperatures are about 1 to 2 times larger than that for method based on zero-field splitting. Thus, the Zeeman level splitting of electronic spins in an external static magnetic field can narrow the linewidth of resonance line and increase the scale factors for temperature measurement.

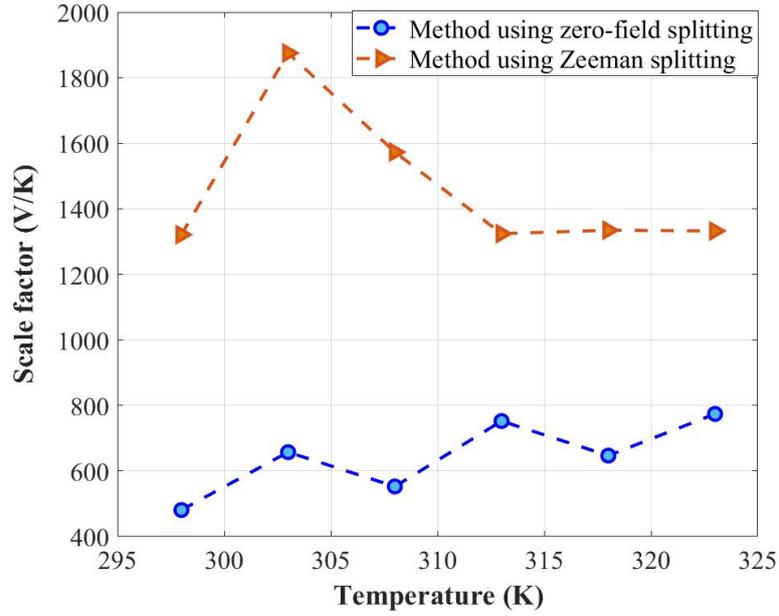

**Figure 4.** The comparison of scale factors for methods based on Zeeman splitting with an applied magnetic field and zero-field splitting without the applied magnetic field.

The sensitivity of temperature measurements can be obtained by power spectral density expansion and the scale factors conversion of raw data[5,31], which are shown in Figure 5. Compared with the method based on ZFS, the low-frequency noise can be significantly suppressed by the method based on Zeeman splitting at the same temperature. We summarize the results of average sensitivity below 1 Hz and 10 Hz in Figure 6 and comparisons of the two methods are presented in Table1. When an external magnetic field is applied to the NV centers, the optimum average sensitivity below 10 Hz for Zeeman splitting method can reach 0.22 K/Hz$^{1/2}$, and that below 1 Hz can reach 0.31 K/Hz$^{1/2}$ at temperature of 303 K. The optimum average sensitivity below 10 Hz for ZFS method is 0.49 K/Hz$^{1/2}$, and that below 1 Hz is 0.66 K/Hz$^{1/2}$ at 313 K. The sensitivity of temperature measurement can be obviously improved by Zeeman splitting method, because the scale factors are increased and the influence of residual magnetic fields, inhomogeneous lattice stress and the mutual interference among different NV axes are simultaneously suppressed.

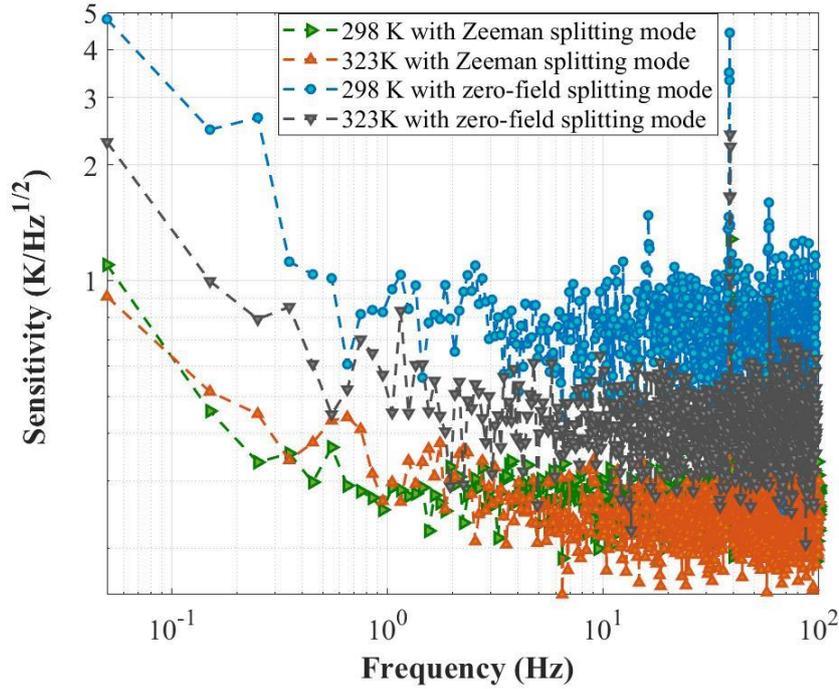

**Figure 5.** Some curves of the sensitivity for temperature measurement under ZFS mode and Zeeman splitting mode.

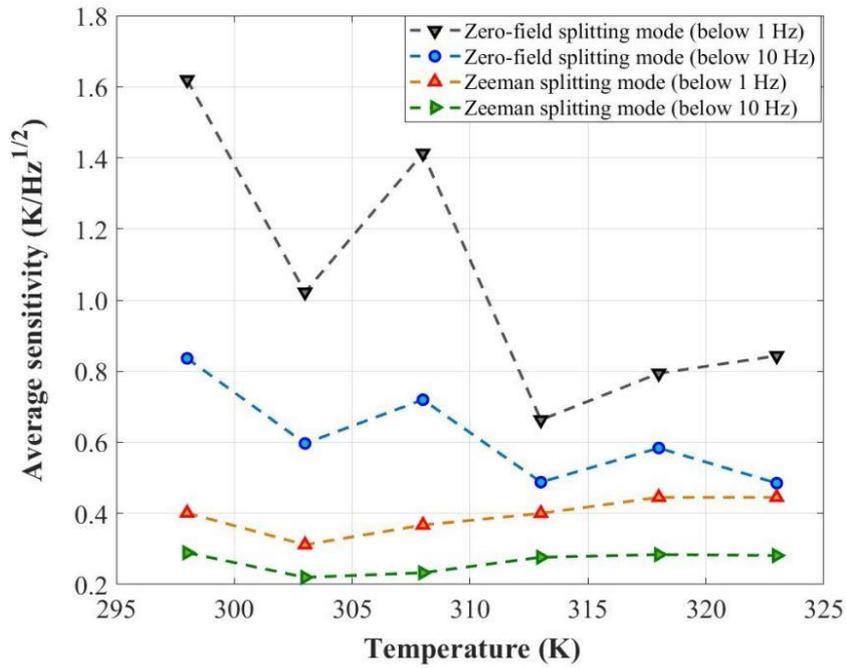

**Figure 6.** Average sensitivity of the NV centers thermometer under different temperatures under ZFS mode and Zeeman splitting mode.

**Table 1.** Comparisons of average temperature measurement sensitivity

| Temperature (K) | Average sensitivity below 1 Hz (K/Hz$^{1/2}$) | | Average sensitivity below 10 Hz (K/Hz$^{1/2}$) | |
| --- | --- | --- | --- | --- |
| | Zeeman splitting | Zero-field splitting | Zeeman splitting | Zero-field splitting |
| 298 | 0.40 | 1.62 | 0.29 | 0.84 |
| 303 | 0.31 | 1.02 | 0.22 | 0.60 |
| 308 | 0.37 | 1.41 | 0.23 | 0.72 |
| 313 | 0.40 | 0.66 | 0.28 | 0.49 |
| 318 | 0.45 | 0.79 | 0.28 | 0.58 |
| 323 | 0.44 | 0.84 | 0.28 | 0.49 |

## 5. CONCLUSION

In conclusion, our study proposed a temperature measurement approach based on the Zeeman splitting of electronic spins using NV centers with multiple-poly crystal directions in diamond. Comparing with the measurement based on ZFS, this method could effectively narrow the linewidth, which has been decreased from 21 MHz to 9 MHz in the experiments. It could also suppress the influence of resonance peaks caused by non-zero $E$, and the fitting errors of the ODMR spectral lines have been reduced. Moreover, we applied a static magnetic field to generate two pairs of Zeeman splitting resonances, and selected the outside pair of resonance peaks generated by a single NV axis to avoid the mutual interference among different NV axes. Furthermore, thermometer was insensitive to magnetic field fluctuations under the Zeeman splitting mode. The accuracy of $D$-$T$ relationship measurement has been significantly improved, and the STD of $D$ values obtained by multiple measurements was reduced by 2 times. The calibration results of coefficient $dD/dT$ is 75.33 kHz/K. The average sensitivity (below 10 Hz) has been improved from 0.49 K/Hz$^{1/2}$ to 0.22 K/Hz$^{1/2}$ in our setup. In the future, the temperature sensing based on Zeeman splitting of NV centers may be beneficial to miniaturization of the system because the magnetic shields are not required and the Zeeman splitting is easy to generate and detect. In addition, simultaneous measurement of temperature and magnetic field may be achieved in this mode.

**AUTHOR INFORMATION**

**Corresponding Authors**

**Xiaojuan Feng** –*National Institute of Metrology, Beijing 100029, China*; https://orcid.org/0000-0002-0612-187X; Email：*fengxj@nim.ac.cn*

**Jintao Zhang** – *National Institute of Metrology, Beijing 100029, China*; https://orcid.org/0000-0002-5204-3403; Email：*zhangjint@nim.ac.cn*

**Authors**

**Li Xing** –*National Institute of Metrology, Beijing 100029, China*; https://orcid.org/0000-0003-3300-8132; Email：*xingli@nim.ac.cn*

**Zheng Wang** – *Department of Precision Instrument*, *Tsinghua University*, *Beijing 100084*, *China*; https://orcid.org/0000-0002-4385-5653; Email：*zheng-wa18@mails.tsinghua.edu.cn*

**Funding**

This work was supported by the China Postdoctoral Science Foundation (No. 2021M703049) and the Fundamental Research Program of National Institute of Metrology, China (No. AKYZD1904-2).

**Notes**

The authors declare no competing financial interest.